# Scalable photonic crystal waveguides with 2 dB component loss

Yahui Xiao[1†], Feifan Wang[1†], Dun Mao[1†], Thomas Kananen[1], Tiantian Li[1], Hwaseob Lee[1], Zi Wang[1], and Tingyi Gu[1*]

[1] Department of Electrical and Computer Engineering, University of Delaware, Newark, DE 19716, USA

**ABSTRACT.** Periodic or gradient subwavelength structures are basic configurations of photonic crystals and metamaterials. The measured linear losses of those nanophotonic devices are well-beyond theoretical predictions. Nanofabrication related geometric inhomogeneity is considered as the primary cause of the deleterious performance. The deep-UV photolithography in CMOS foundry is a large-scale parallel processing, which can significantly suppress the random offsets and thus the optical linear loss. Here we demonstrate ultra-low loss photonic crystal waveguides with a multi-project wafer run through AIM photonics and post-processing. For sub-millimeter long photonic crystal waveguides, 2 dB total loss and 40 dB extinction ratio are observed across dies.

## 1. INTRODUCTION

Suspended integrated photonic structures play a critical role in the studies of optomechanics [ma], MEMS based electrical and optical switch [se] and high-speed optoelectronic transceivers [lm]. Air gap with lower refractive index reduces evanescent coupling from the guided modes to substrate, and thus improve the resonator quality factor and reduces propagation loss. In ultrafast MZM modulators, the removed substrate reduces the transmission loss in electrodes [lm]. Two dimensional photonic crystals(PhCs) also utilizes suspended photonic structures. Line-defect waveguides (WGs) are compact implementation of slow-light optical delay lines [2,3], directional couplers [4,5], and active components with enhanced light-matter interactions [3,6-9, tan].

Nanomanufacturing of the PhC structures were firstly investigated in foundries of IMAC [10] and IME [19]. The subwavelength periodic structures are topologically different from channel waveguides. The proximity effects require adjustment in dosage for PhC devices [10,19]. Fabrication induced geometric offsets shift the operation wavelength of PhC WGs, which needs to be calibrated in each process line. The geometric inhomogeity introduces optical 'defect states' in the PhC bandgap structure. If the 'defects states' spectrally overlap with PhC WG modes, the photon leakage to the the PhC increase spropagation loss. The defect states out of the WG wavelength range tunneling states, and reduce the extinction ratio (ER). ER is the transmission contrast between the light within and out of the WG wavelength. Low propagation loss around 10 dB/cm has been reported in PhC WG [19,21,5]. While the propagation loss calibration requires the comparison between a set of WG structures, the ER can be easily identified even with one transmission spectra, even in devices with large coupling losses. The ebeam lithography fabricated PhC WGs have ER around 20-25 dB [13,22,20,5]. The DUV manufactured PhC WGs are found to have higher ER ~40 dB [19,21], which indicates improved geometric uniformity provided by DUV [14].

The increasing technical readiness level of silicon photonics shift the research focus from the device to circuits level [1]. From the circuit-design perspective of view, an individual device's 'component loss', including input/output channel WG coupling loss and the total propagation loss, is of paramount importance in the device's performance matrix [se]. The component loss of sub-mm long PhC WGs is reported to be around 10 dB in most works [19,13,21], and can be further improved to be 4-5 dB [5, 23]. Here we demonstrate low-loss PhC WGs manufactured by AIM photonics [aim]. The total loss, including both insertion loss and the propagation loss over 0.2-mm-long PhC WG, is measured to be around 2 (±1) dB in devices across different dies. The extinction ratio between the pass and stop band is measured to be around 40 (±4) dB. The minimal tunneling photons in the stopband indicate excellent geometry homogeneity. Those devices are manufactured through a multi-project wafer run on a 300-mm wafer and post-processed in the local cleanroom for the undercut. With the carefully designed undercut profile, the two-stage three-dimensional coupler between channel WGs on buried oxide and suspended PhC membrane shows less than 1 dB insertion loss. Given similar WG length (0.2-1 mm), the DUV fabricated devices exhibit higher extinction ratio and lower linear loss. The reduced linear loss is critical for PhC applications in large-scale PICs for optical signal processing and interconnects.

## 2. RESULTS AND DISCUSSION

The lattice constant of the PhC air holes is fixed as 440 nm, with the radii vary from 135 nm to 170 nm by an increment of 5 nm for the geometric offsets. After the pattern layout passes the design ruler check (DRC) for foundry compatibility, the PhC devices defined by DUV photolithography are manufactured as multi-project wafer (MPW) passive run AIM Photonics.

Arrays of PhC devices are manufactured on a 300 mm silicon-on-insulator (SOI) wafer with a 220 nm-thick device layer. The lattice constant and radius of the PhC are more than the critical dimension of the foundry. A thick PECVD silicon oxide film protects the silicon layer. Post-processing selectively etches the substrate and superstrate of the PhC area. Firstly, this top oxide is removed by wet etching with buffered oxide etching (BOE) 1:6 at a rate of about 160 nm/min.

A photoresist (AZ1512) is then spin-coated onto the sample, followed by laser writing to open a window above the PhC region. The window shrinks to the inner area to keep the edges of PhCs untouched by the liquid. The zig-zag shape helps to keep the PhC stable, even when it is suspended. After the laser writing, the sample is dipped into BOE 1:6 to do the undercut. The sample is kept in BOE etchant for 30 minutes. Finally, the remaining resist is stripped. Fig. 1(b) shows the optical images of the device near the PhC WG-channel WG interface for steps. After the wet etching, the underlying oxide in the polygon area is removed. After stripping off the resist, the suspended area changes its structural color from pink to green. The green color indicates the thermal oxide under the PhC is removed. Measurements show that the etched area is larger than the defined area, about 1 micrometer in the plane. As the etching is isotropic, we estimate 1-micron thick thermal oxide is removed under the PhC plane.

Fig. 1(a) represents the side-view design of the two-stage channel WG – PhC WG coupler, where the input light travels from the low loss channel WG to the supported PhC WG and then to the suspended PhC WG region. Post-processing of undercut to increase the vertical index contrast for light confined by total internal reflection. The overall loss is measured to be less than 1 dB. Fig. 1(b) shows the top view of the optical image on the coupling region, where the dashed curve represents the interface between the supported and the suspended region. The green color inside the zig-zag window indicates the suspended region after the undercut. Fig. 1(c) shows the perspective view of the area near the PhC WG and the edge of the window. The SEM images show little oxide residue (or resist polymer residue) and nonuniformity of the periodic PhC structure. The close view of the shadow box at the interface between supported and suspended PhC WG after post-processing is shown in Fig. 1(d). Inside the window defined on the photoresist, the thermal oxide under the PhC layer is removed to reduce light coupling to the substrate. In comparison, some residuals are left outside the window to improve the mechanical stability of the PhC structure.

Fig. 2(a) illustrates the optical microscope image of the PhC WGs array after undercut. We designed the W1 PhC WG with a targeting wavelength of around 1550 nm. The lattice constant of the PhC air holes is fixed as 440 nm, with the radii varying from 135 to 170 nm by an increment of 5 nm for the geometric offsets. Different radii show a different color according to the optical reflection. The post-processing by wet etching undercut for a certain PhC WG region to increase the vertical refractive index contrast. Fig. 2(b) shows an SEM image of PhC WG on the coupling region. The designed channel WG is 500 nm, while the channel WG from SEM measurement shows 455 nm, possibly relating to SEM self-distortion and fabrication exposure dosage. We measured lattice constant a and radius r of PhC WG arrays from two chiplets. Fig. 2(c) and (d) represent the comparison between SEM measurement and design. There are double measurements of chiplet I because of different air hole shapes from short (orange dot) and long (orange square) exposures under SEM. Chiplet II (open window) in the blue dot represents the PhC WGs region with only top oxide removed. Different air hole sizes of the same device from chiplet I and II may reflect the various positions in a wafer. The chiplet in the center will have more exposure than the one at the edge during the lithography process.

To characterize the loss of the PhC device, we measured broadband transmission spectra and compared the results across the chiplets. A continuous-wave laser tunable from 1480 to 1580 nm is directed to the input grating coupler by single-mode fibers. The signal from the output grating coupler is connected to a photodetector. We tested the PhC WG transmission spectra before and after the undercut. The transmission spectra of PhC WG with the design hole radius of 135 nm and 140 nm in chiplet I are shown in Fig. 3(a) and (b), respectively. Before the undercut (gray curves), the total loss of the PhC device is larger than 10 dB, and there is no clear band-edge between 1480-1580 nm. A clear and sharp band-edge appears after the undercut (orange curves) with an extinction ratio over 40 dB. The flat curve before the band edge indicates an insertion loss around 2 dB for this 0.2-mm-long PhC WG. The corresponding photonic band diagram of the design radius r=135 nm (measured as a=430 nm and r=122 nm) and r=140 nm (measured as a=430 nm and r=122 nm) PhC devices shown in Fig. 3(c) and (d) indicate the guided mode in PhC WG overlaps with the vertical dot lines (tunable laser range) around a normalized frequency of 0.296, matching the high transmission part after fully undercut in Fig. 3(a) and (b). While the PhC device before the undercut is indicated in gray dashed line ($n_{sub}$) in Fig. 3(c) and (d) has no guided mode in the frequency range of the tunable laser, matching the lossy curves (gray) in the transmission spectra.

Since the total loss includes both insertion loss and propagation loss, we measured the fiber-to-fiber transmission spectra of PhC WGs from two different chiplets with grating coupler loss (gray curve) to distinguish the component loss shown in Fig.4(a) and (b). The inset schematics show the grating coupler and PhC WG. The ripples of coupling efficiency (gray curve) are mainly from Fabry-Perot oscillation. The radius of working devices from chiplet I are r=135 nm (blue curve) and 140 nm (blue curve), respectively. Similarly, the radius of working devices from chiplet II are r=150 nm (blue curve) and 155 nm (orange curve). The difference in the working radius between chiplet I and chiplet II is the minute post-processing time difference during the wet etching. In this case, chiplet I has a slightly longer time sinking in the BOE solution than chiplet II, which caused the working radius to shift to a smaller size.

Fig. 4(c) compares the reported performance of PhC WG fabricated by DUV and E-beam lithography. Given similar WG lengths (0.2-1 mm), the PhC devices presented in this work exhibit a higher extinction ratio and lower linear loss. The reduced linear loss is critical for PhC applications in large-scale PICs for optical signal processing and interconnects.

**DISCUSSION**

In summary, we demonstrate the first PhC devices through AIM MPW run, with ultra-low component loss (~2 dB) for large-scale integration. The high extinction ratio near the PhC WG passband edge indicates excellent geometric homogeneity, with minimal tunneling photons through the defect states in the PhC bandgap.

**METHODS**

Layout Preparation: The W1 PhC waveguide is designed by Finite-difference time-domain method [24], with operation wavelength in telecommunication C band. With a fixed lattice constant of 440 nm, the geometric offsets are introduced into the void radii (from 135 nm to 170 nm with an increment of 5 nm). The width of the channel WGs is fixed to be 500 nm.

Fabrication: Those deep UV photolithography defined PhC devices are manufactured as an MPW passive run at AIM Photonics. The PhC device in Fig. 1b is fabricated at university cleanroom. The PhC layout was firstly defined in CSAR 6200.09 positive resist layer by using a Vistec EBPG5200 electron beam lithography system with 100kV acceleration voltage, followed by resist development and single-step dry etch procedures

Measurement: We measured the transmission spectrum and insertion loss of the PhC WG through the devices by a coupled fiber system. A continuous-wave laser tunable from 1480 nm to 1580 nm (with a spectral resolution of 10 pm) is directed to the input grating coupler by single-mode fibers. The signal from the output grating coupler is connected to a photodetector.

**ASSOCIATED CONTENT**

**Correspondent Author**

Tingyi Gu – Electrical and Computer Engineering, University of Delaware, Delaware, United States; *E-mail: tingyigu@udel.edu

**Author Contributions**

F. W., Y. X., and D. M. performed postprocessing and measurements. T. K. and T. L. prepared the TAPEout. F. W., Y. X., D. M., and T. G. analyzed the device results. Y. X., F. W., and T.G. wrote the manuscript, and all authors discussed the results and commented on the manuscript.

[†]Y. X., F. W., and D. M. contributed equally to this work.

**Notes**

The manufacturing process information is not available according to the non-disclosure agreement between the authors' institution and AIM Photonics.

**ACKNOWLEDGMENTS**

   The devices are fabricated by AIM Photonics and post-processed at the University of Delaware Nanofabrication Facility. This work is sponsored by Air Force Office of Scientific Research Young Investigator Award via Award Number (FA9550-18-1-0300) and National Aeronautics and Space Administration Early Career Faculty Award (80NSSC17K0526).

**FIGURES:**

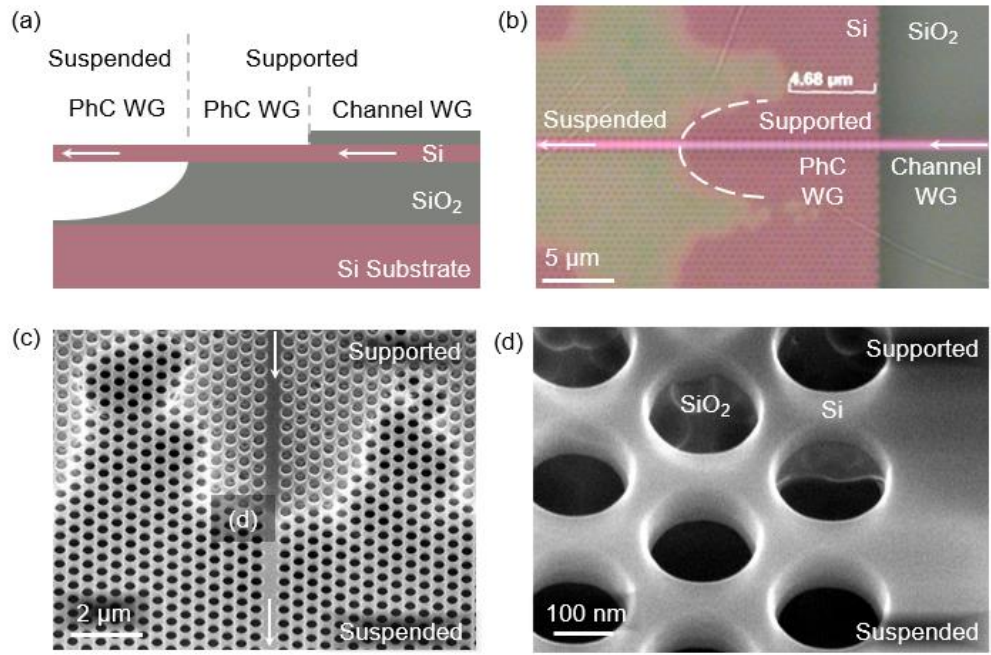

Fig. 1. Low loss three-dimensional coupler between a channel WG and a PhC WG. (a) Side-view of the channel WG – PhC WG coupler. (b) Top view of the optical microscope image on the coupling region. (c) Perspective view of the SEM image. (d) Close view of the interface between supported and suspended PhC WG after post-processing.

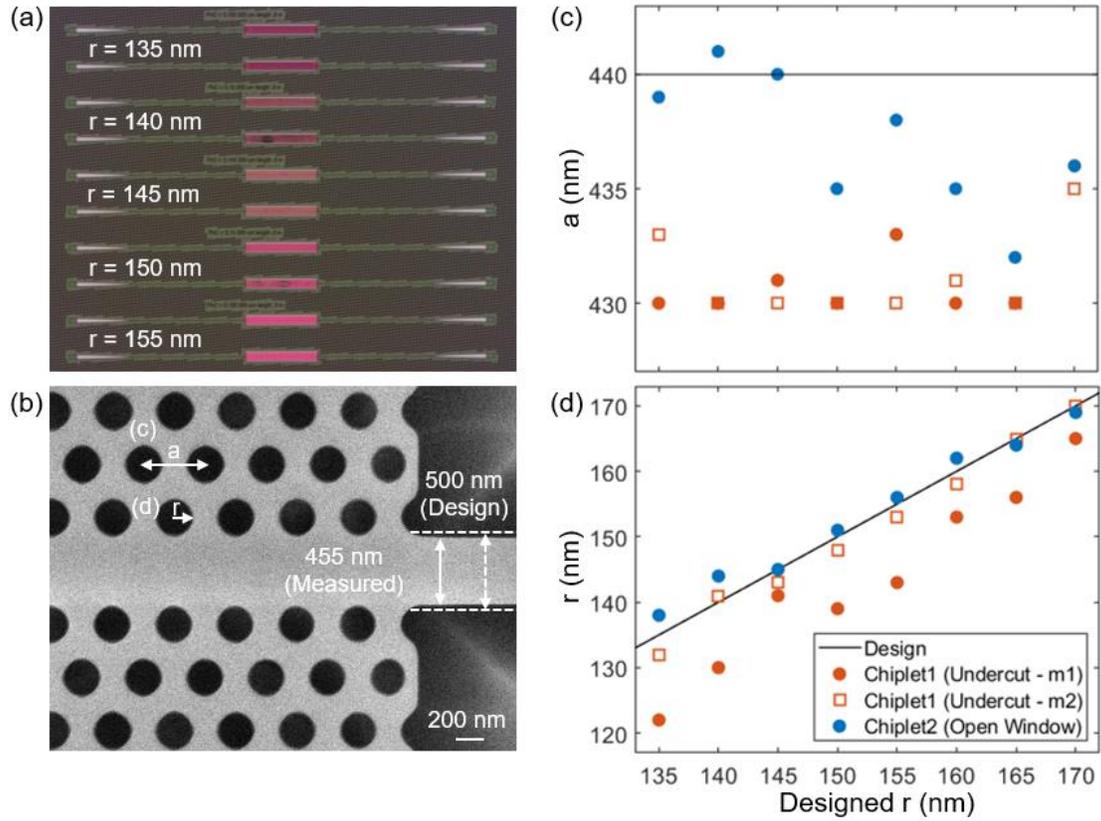

Fig. 2. Measurement of PhC parameters from different chiplet devices. (a) Optical microscope image of PhC WGs (after undercut) with radii increasing from r=135 nm to r=155 nm with 5nm increment. (b) SEM image of PhC WG on the coupling region. Lattice constant a and radius r are measured as the arrows show. (c) and (d) are the measurements of a and r from SEM images in Fig. 2(b) versus the designed r. The black line represents the designed parameters.

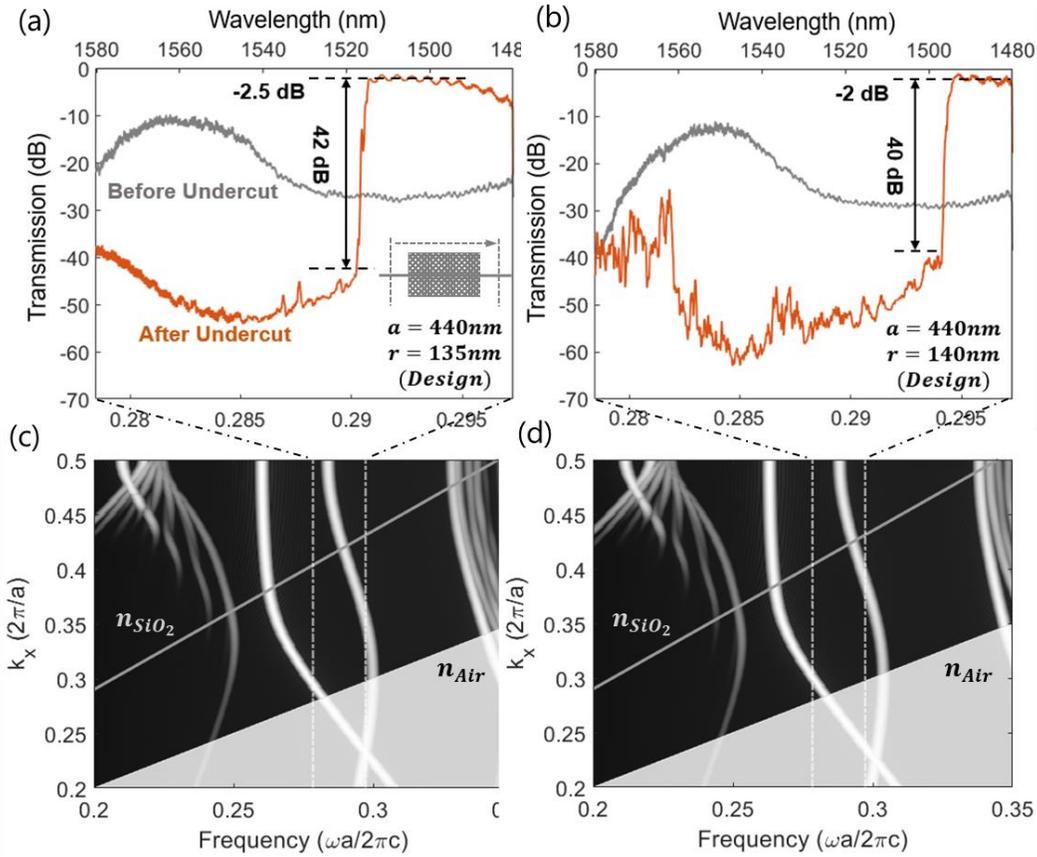

Fig. 3. Measured total loss of manufactured PhC WGs across chiplets (grating coupler loss excluded). The PhC lattice constant is at 440 nm. (a) and (b) are the radius of 135 nm and 140 nm PhC devices before (gray) and after undercut (orange); (c) and (d) are the corresponding photonic band diagram of the PhC WG for TE-like modes. The vertical dot lines show the tunable laser range in (a) and (b). The dashed gray line indicates the light line of the substrate (before the undercut). [The band diagram calculation is updated compared to the publication in PTL2022]

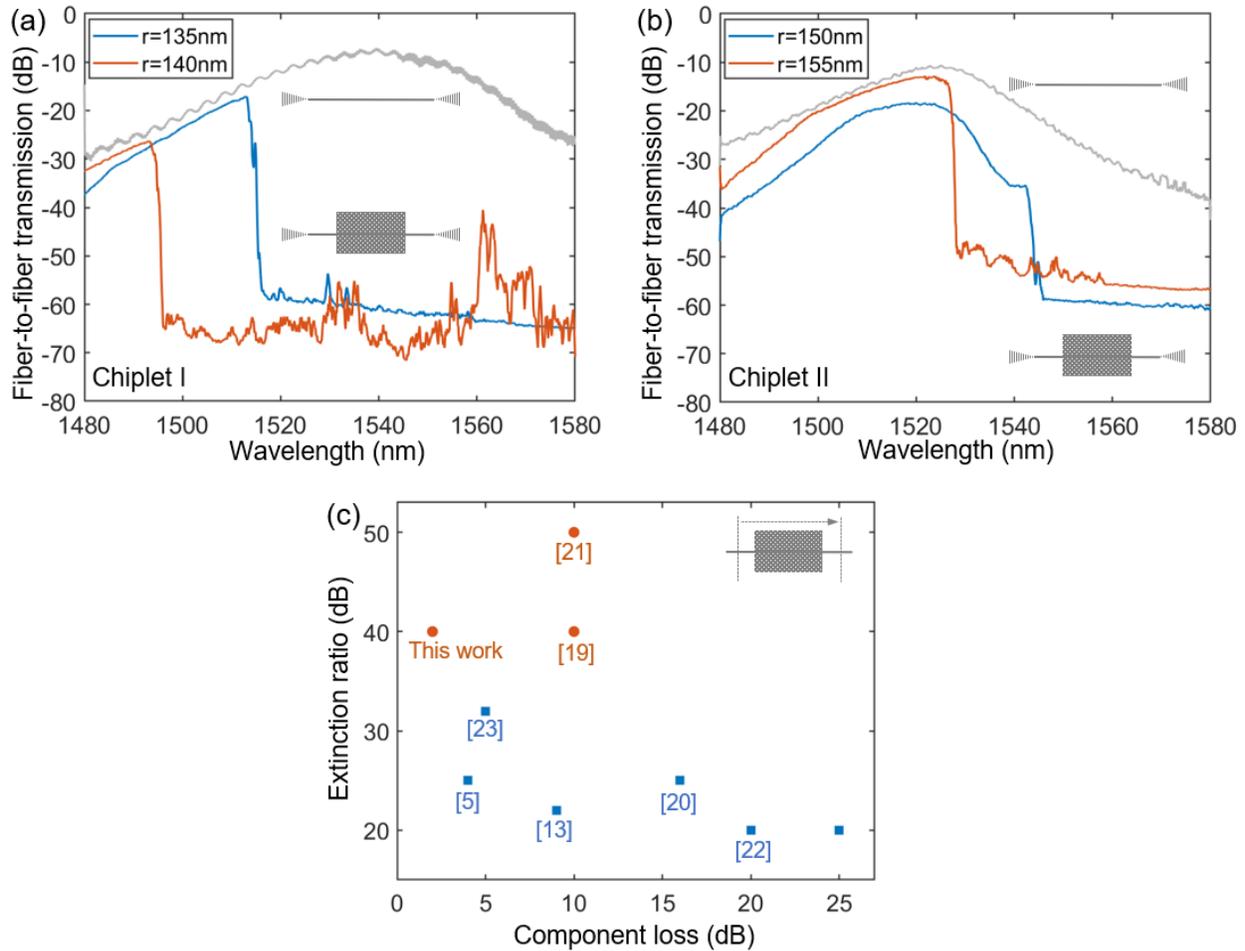

Fig. 4.: Measured fiber-to-fiber transmission spectra of manufactured PhC WGs (grating coupler loss included) with the corresponding coupling loss (gray). The PhC lattice constant a=440 nm. (a) Chiplet I: transmission spectra of the radius r=135 nm (blue) and r=140 nm (orange) PhC device after undercut; (b) Chiplet II: transmission spectra of the radius r=150 nm (blue) and r=155 nm (orange) PhC device after undercut; (c) Linear response comparison for mm-long PhC WGs [5,13,19-23]. The red dots represent PhC WGs by DUV lithography, and the blue squares represent E-beam lithography.